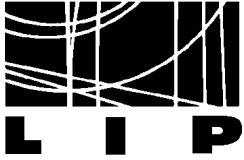

# Quality Control of GEM Detectors Using Scintillation Techniques

F.A.F. Fraga[1,2,a], S.T.G. Fetal[2], R. Ferreira Marques[1,2] and A.J.P.L Policarpo[1,2]

[1] *LIP - Laboratório de Instrumentação e Física Experimental de Partículas - Delegação de Coimbra, 3004-516 Coimbra, Portugal*
[2] *Departamento de Física, Universidade de Coimbra, 3004-516 Coimbra, Portugal*

**Abstract**

Non destructive quality control of microstructures at the manufacturing stage is an important issue in the foreseen use of huge numbers of such gaseous detectors in the future high luminosity colliders. In this work we report on the use of the scintillation light emitted by the avalanches in GEM channels for checking defects in the foils. The test system is described and data on the relative efficiency of several gaseous mixtures are presented. The foil images obtained with a low-noise CCD system are analysed and compared with the optical images obtained with an industrial inspection system of high magnification.

The validity of this test method is established and possible extensions of its use are discussed.

## Introduction

The development of gaseous detectors using microstructures is under very active research, many new designs turning up recently. The results obtained up to now with the GEM (gas electron multiplier) [1] are very promising (amplification factor up to several thousands, count rate up to $10^5$ counts mm$^{-2}$s$^{-1}$) and, due to its simplicity and ease of manufacture, it is now available on the market. It can be expected that large quantities of these foils will be produced and their quality control at the manufacture stage will be an important issue in the assembly of future detectors. Currently, the foils are randomly inspected immediately after etching by visual observation under magnification and some simple tests based on resistance measurement are done before they are accepted or rejected [2].

Although some systems for test of microstructures with anode strips have been presented, they basically rely on continuity and/or insulation measurements. However, such tests, that supply valuable although limited information about the operating condition of the microstructure, can not be performed with surface type devices such as the GEM.

Recently, following the observation that microstrip plates can be operated in pure noble gases and that a large light emission is then observed, we presented some work on the use of this scintillation light for quality control of microstructures and considered the possibility of exploiting it to test GEM foils using an optical system based on CCDs [3].

The aim of this paper is to present experimental results from such a system and to compare them with those obtained by visual inspection of the foils under high magnification. We establish this method as an worth considering tool for the prototyping and development of microstructures and, in particular, suited for application when it is matter of manufacturing large arrays of microstructure detectors.

## Experimental system

An existing stainless steel test chamber was adapted for carrying these experiments. A schematic cross-section of the chamber is shown in fig. 1. The entrance window was a 50 µm aluminium sheet and a 10×10 cm$^2$ GEM foil was used for the measurements, although only a small zone ( ≈ 2.5 cm$^2$) of it was irradiated, due to geometrical limitations around the entrance window. The GEM, from the CERN RDD Group, was manufactured from a 50 µm thickness copper coated kapton foil by a chemical etching process that produces holes with a double conical shape (standard shape) [4]. The diameters of metal and kapton holes were 80 µm and 55 µm, respectively. The pitch was 140 µm and the optical transparency ≈12%. The GEM front side was grounded, the back one was operated at negative voltage. A drift plane was placed 4.5 mm before the GEM, whereas the collecting wires, with a pitch of 2 mm, were placed 6.5 mm away from the GEM to reduce the shadowing effects on the CCD images. The light window, made of glass, was 6 cm in diameter.

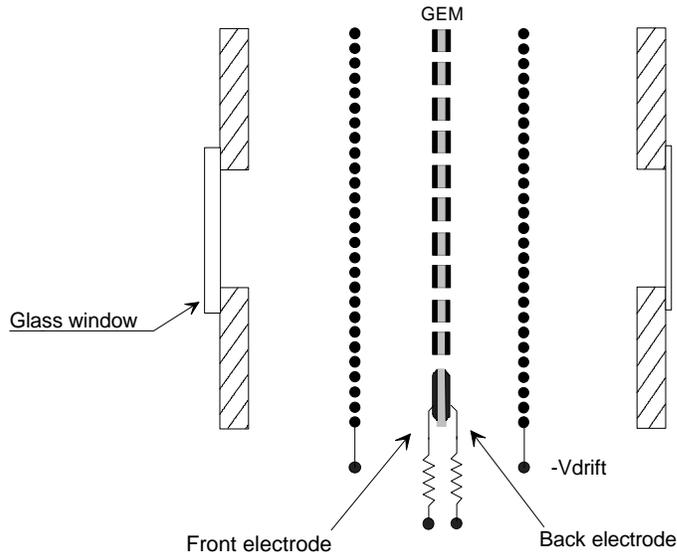

Fig. 1 Cross-section view of the detector chamber used in this work.

A diagram of the complete system is shown in fig. 2. The primary electrons produced by the X-ray photons in the conversion region drift towards the GEM holes. The grid currents were monitored with the high voltage power supply ammeters with a resolution of nA and the outer GEM grid was connected to the ground through a Keithley picoammeter, allowing for precision measurement of the collected electrons current. Due to the large distance between the GEM and the collecting grid and the chosen operating voltages, all the electrons produced in the multiplication zone were collected at the outer electrode of the GEM, and the ratio between primary current and this current was used to calculate the GEM gains. A Quantix 1400 camera, manufactured by Photometrics Ltd, was used to readout the light emitted from the GEM. It uses a Peltier cooled, low noise CCD, KAF 1400 from Kodak, with $1317 \times 1035$ pixels of $6.8 \times 6.8$ μm$^2$ and the spectral response is shown in fig. 3. The camera was operated with the LVIEW software [5] and further image analysis was carried with the GW package [6]. A standard 50 mm photographic lens was used and the camera placed at the minimum allowable focusing distance, about 30 cm away from the GEM foil.

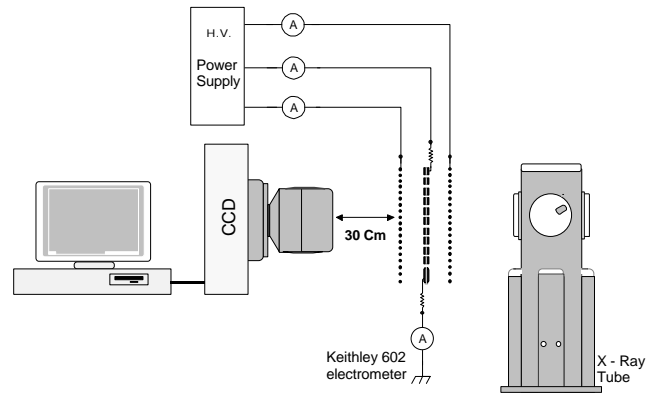

Fig. 2 Schematic drawing of the experimental system.

The chamber was irradiated by an X-ray generator with a molybdenum tube. As this tube was operated at low voltages, typically around 10 kV, the main bremsstrahlung spectrum was peaking around 8 keV. The emission rate could be controlled either by tube current adjustment or placing absorbers in the beam path, or both.

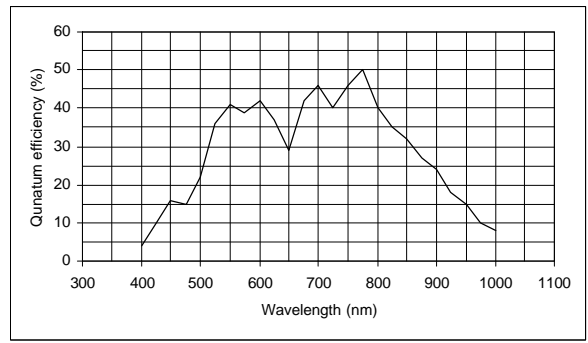

Fig. 3 Quantum efficiency of the Quantix 1400 camera versus wavelengh (data supplied by the manufacturer).

The gases used, all high purity research grade, were supplied to the chamber through stainless steel pipes without any additional purification. The detector was operated in a closed system mode, the gas being kept for periods of up to five days.

**Experimental results**

Fig. 4 shows an image of the scintillation light produced at the GEM foil. The CCD image was obtained by irradiation of a small circular area (~16mm diameter) of the detector during 100 s. The gaseous mixture was Ar-5% $CO_2$. The drift and collection grids potential were -800 and +800V, respectively, and the applied voltage between the GEM electrodes (VGEM) was 345 V. The drift potential was set at the maximum voltage that avoids collection of primary charges by the GEM back copper electrode, for gain calculations the GEM electrical transparency was considered close to 100%. The current of the X ray tube was 2 mA, corresponding to a detected flux of $4\times10^{-4}$ counts $mm^{-4} s^{-1}$.

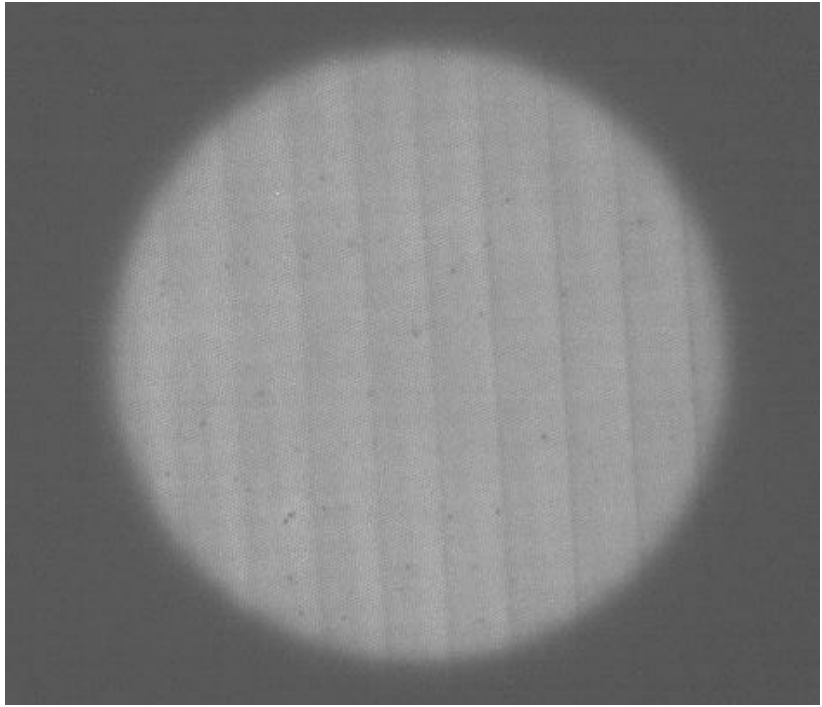

Fig. 4 Image of the scintillation light produced by the GEM avalanches in the irradiated zone. The X-ray detected flux was $4\times10^4$ counts $mm^{-2} s^{-1}$ and collection time was 100 s.

The light emitted from the GEM holes is clearly seen. The equally spaced vertical defocused shadows are due to the collecting wires ($g_2$) and the square patterned background, visible although out of focus, is the image of the drift grid seen through the optically transparent GEM. Disregarding these global regular variations of light intensity, some local variation of the intensity of the light associated with each GEM hole can be seen. Some holes are completely dark, and a few others have light emissions noticeable different from their neighbours, suggesting that they had different electrical field value and configuration.

Although simple arguments point towards the light emission being produced in the GEM holes due to the higher value of the electric field inside the channel, one can not exclude that some light be produced when the electrons drift towards the collecting grid. The electrons due to the avalanches inside the GEM holes can be collected by either the front GEM electrode or the collecting wires, depending on the value of the electric field in the collection region. In our configuration, all electrons are collected by the GEM when a negative voltage is applied to the collecting grid. No variation of the GEM illumination was noticed when the potential of the collecting grid was varied between +2500V and -800V, showing evidence that all the light seen by the CCD was produced inside the GEM holes and the local variation of intensity could be associated to each of them.

In order to perform the tests at a safe low $V_{GEM}$ voltage, reduce the exposition time and improve the signal to noise ratio of the images, the gas mixture should be selected in a way that the light emission efficiency is maximal over the spectral range of the CCD. We took measurements of the amount of the light collected by the CCD with different mixtures. Ar, Xe, Ne and Kr mixtures were tested, but Ar mixtures yielded higher light emission.

The variation of the light over current ratio versus $V_{GEM}$ for several quencher concentrations of the Ar-$CO_2$ gas filling is shown in fig. 5. Although the luminosity of the pure gas was higher than with quencher, the mixture was unstable, showing some dependence on the impurities content, as evidenced by the time evolution of the measurements. It should be stressed, however, that the charge gain was stable under those conditions and only the light emission was affected by this very small impurity content.

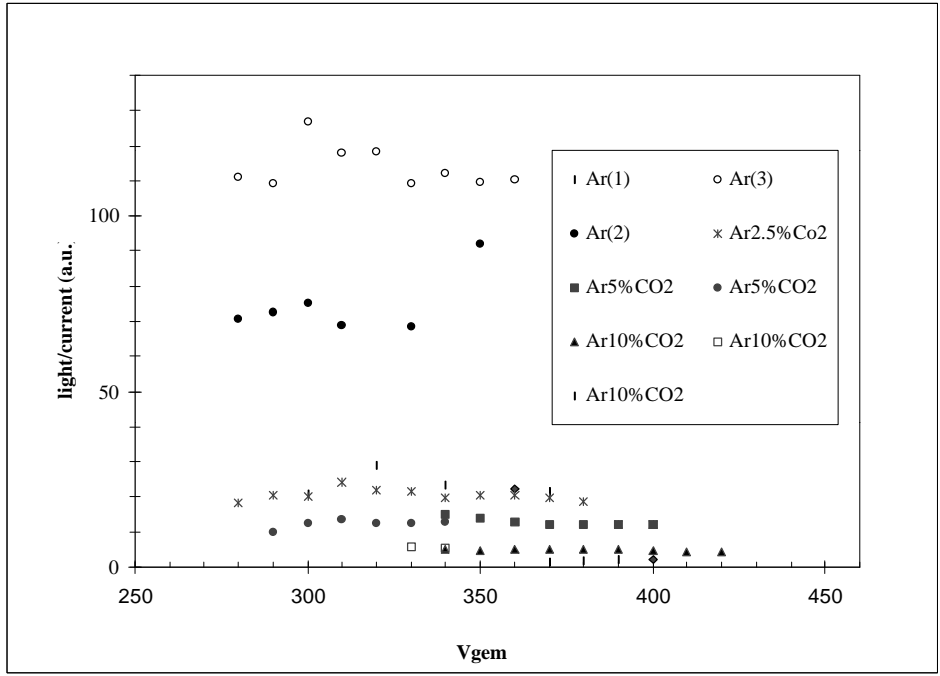

Fig. 5 Ratio of emitted light over electron current versus $V_{GEM}$ for several quencher concentrations of the Ar-$CO_2$ mixture. Ar(1) is pure argon several days after being filled, Ar(2) and Ar(3) were measured immediately after filling the detector.

The results obtained with Ar-$CO_2$ mixtures show that the light emission is reduced by the addition of quencher, although the emission becomes foreseeable and stable in time. For a given mixture the number of photons per electron emitted is almost constant along a large plateau, showing a small variation with $V_{GEM}$.

Measurements of the light emission versus count rate were also performed. The dependence of the mean amplitude of the collected light versus X-ray count rate is shown in fig. 6. The results of light emission show the same drop that can be seen in similar plots of charge gain of the GEM versus count rate. Although we have not performed precise measurements of the dependence of charge gain on counting rate, the similarity of those curves suggests that this is also due to space charge effects which disturb the multiplication in the GEM channels. This explanation was further backed by the observation that the ratio of emitted light divided by electron current was almost independent of count rate.

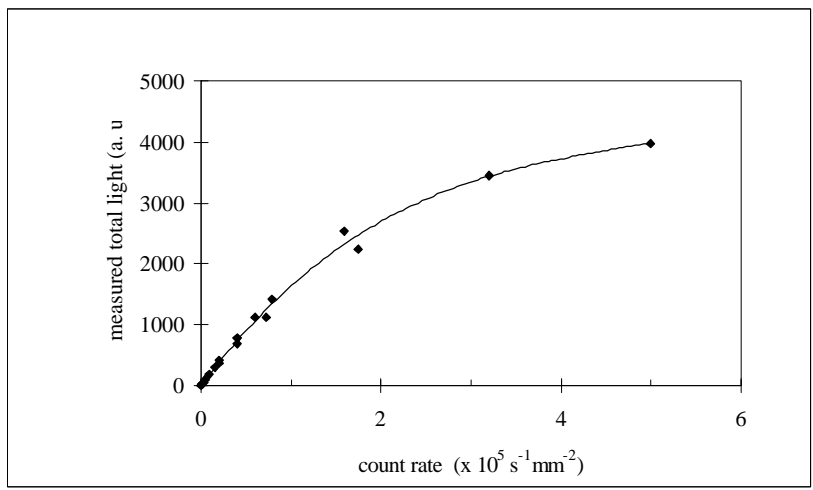

Fig. 6 Light emission versus count rate. Although the ratio of emitted light over electron current is almost independent of count rate, the light is limited by the reduction of GEM gain at high count rates.

Fig. 7 a) and b) show two images of the same small area of a GEM operated with an Ar-5%$CO_2$ mixture. The first was taken immediately upon assembly of the detector and the second one week later, after the GEM had suffered some abuse. In both figures some local variations of the light emitted by individual channels are clearly visible. It can also be seen that more non illuminated channels are present in b) than in a), and that a few non illuminated channels in a) seem to have recovered in b).

After disassembly of the chamber, the GEM foil was observed using a high magnification industrial optical inspection system. Views from both sides of the zone marked with an arrow in fig. 7 are shown in fig. 8 a) and b). The two neighbouring channels seen as darker holes in the scintillation image are really clogged holes, i.e., holes on which one side of the copper clad sheet was not etched. These holes show bright centres in fig 8 a) because of the reflection of the light in the copper surface, which also means that the kapton was incompletely removed. The adjacent channel that is less luminous in fig. 7 a) and b) is really a hole with the copper incompletely removed. Although some other defects first seen in the scintillation image were also found in the subsequent optical observation, no visible defects could be associated to several of the dark channels, including all those that recovered between acquisition of figs. 7 a) and 7 b).

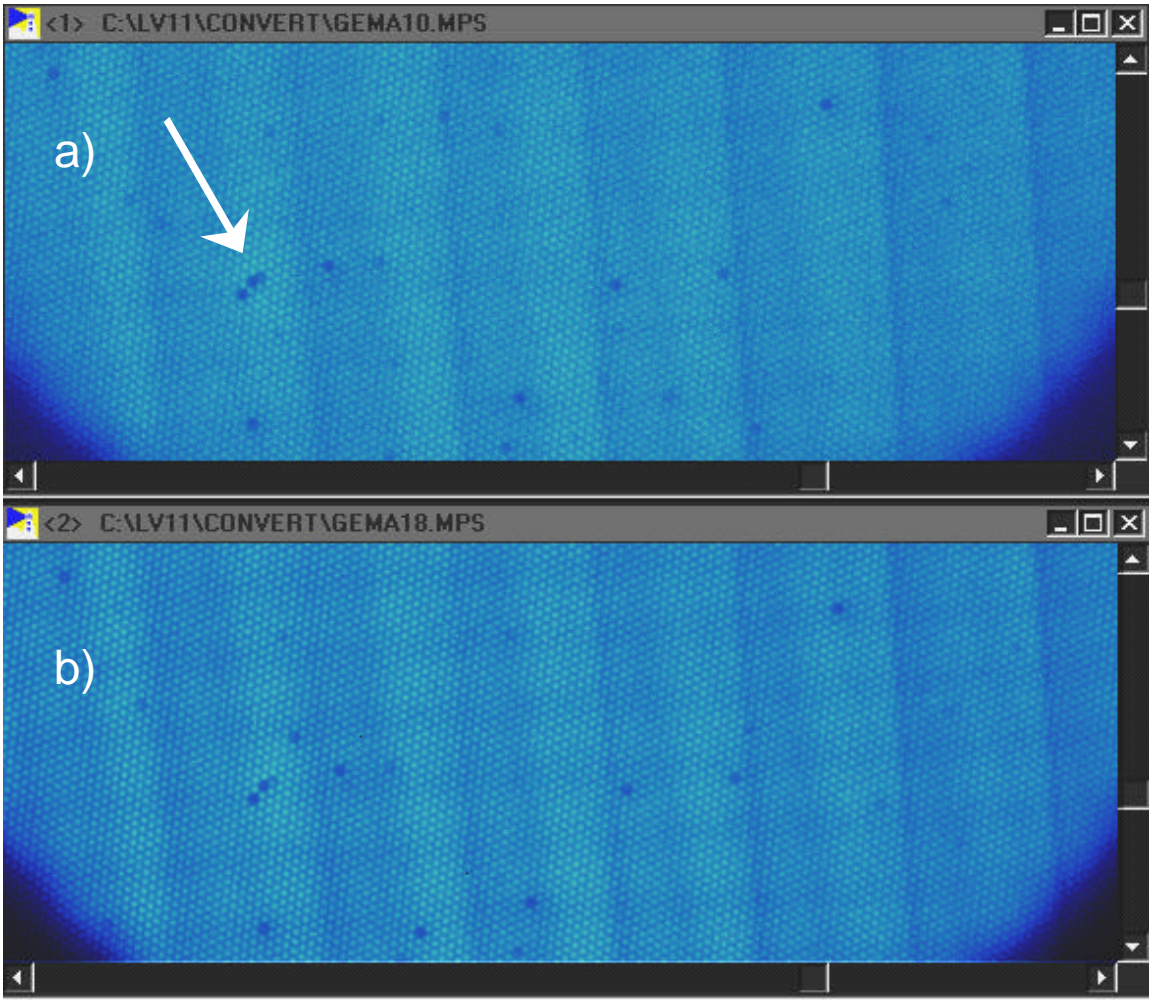

Fig. 7 Scintillation images of the same GEM area taken immediately after assembly (a) and later, after the GEM has been used for some days and suffered some abuse (b).

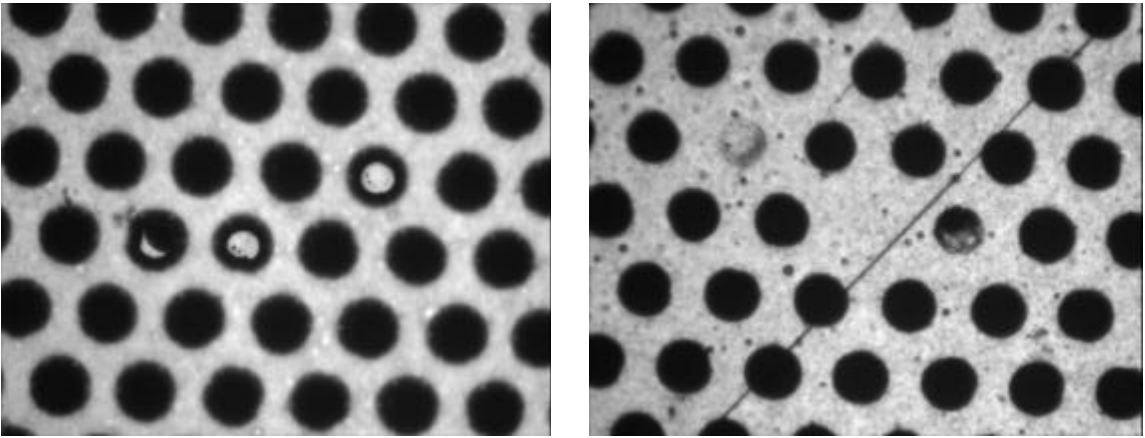

Fig. 8 Images of both sides of the GEM foil area indicated by the arrow in the previous picture, taken with an optical inspection system. The defects in the copper layers and the kapton hole are clearly seen.

During some time one the holes of the GEM exhibited a strong light emission when the polarising voltages were applied, even in the absence of X-ray irradiation, as seen in fig. 9. However, the GEM operated normally and, later on, this channel recovered normal operation. When observed under high magnification, no particularities in its physical aspect have been seen.

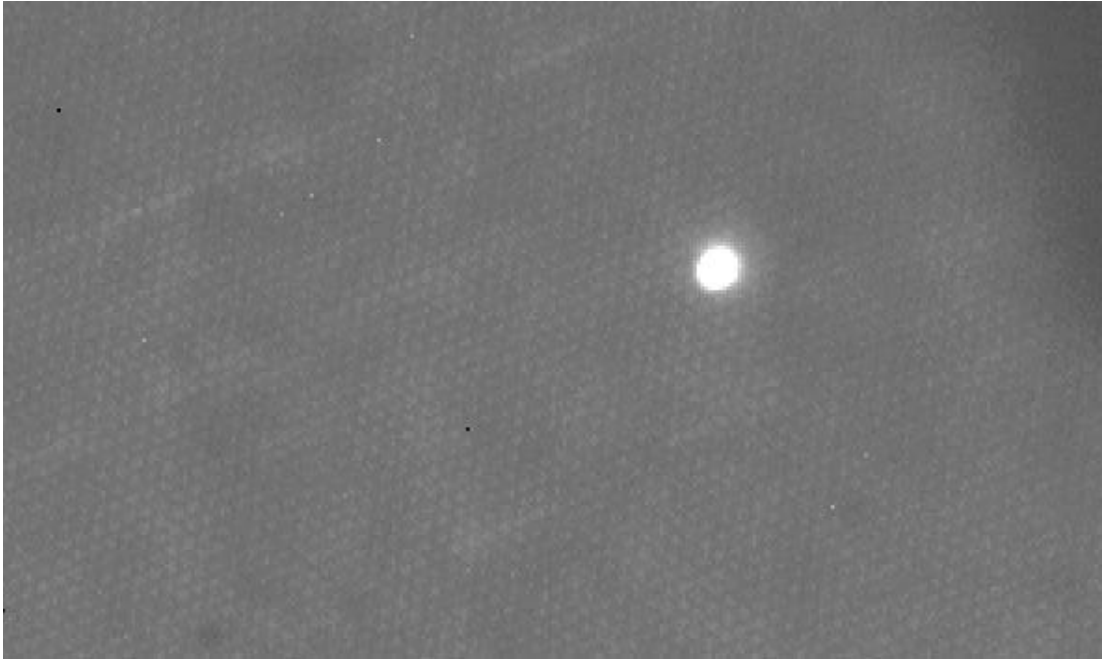

Fig. 9 Temporary self emission of light from one single GEM channel. This emission ceased after some hours of operation and the GEM channel resumed normal behaviour. No vestiges of damage were found by later optical observation of this channel.

Although the detectors were assembled in a clean room, this self recuperation of some GEM defects could suggest that they were due to small particles of dust attached to the foil that later on, due to the normal operation of the detector, were removed from its surface. Some of these defects could also be due to residues left on the kapton surface during the manufacture, which disappear upon operating the detector.

## Conclusions

We have shown that the visible light emitted by the GEM avalanches can be successfully exploited for quality control of the foils, checking their global uniformity and identifying local defects. This technique essentially is sensitive to electric field configuration and then is a priori more adequate for testing purposes than simple optical inspection. Some defects that show up in the scintillation image are not visible under the microscope. X-rays have been used as a suitable electron source; of course other scintillation techniques may turn out to be more convenient.  Image processing software tools must of course be used to take profit of this method. We are currently developing a scanning device that will allow  to test the whole area of a GEM foil.

The use of this method as a tool for microstructure research and development should also be considered.

## Acknowledgements

This work was supported by the contract CERN/P/FIS/1198/98 with the portuguese FCT. Thanks are due to F. Sauli (CERN GDD) who made available the GEM foils used in this work.